\documentclass{aa}  
\usepackage{graphicx}
\usepackage{amsmath}	
\usepackage{amssymb}	
\usepackage{relsize}
\usepackage{txfonts}
\usepackage[colorlinks=true,linkcolor=blue,allcolors=blue]{hyperref}%

\newcommand{\Msun}{M$_\odot$}

\newcommand{\rcut}{\ensuremath{r_\mathrm{cut}}}
\newcommand{\rmax}{\ensuremath{r_\mathrm{max}}}

\usepackage[colorlinks=true,linkcolor=blue,allcolors=blue]{hyperref}%

\usepackage{ulem}

\begin{document}

\title{Distinguishing the formation paths of massive compact ETGs through their internal dynamical structures }

   \author{Ling Zhu$^1$\thanks{Corr author: lzhu@shao.ac.cn}, Ana L. Chies-Santos$^2$, Micheli Trindade Moura$^2$, Hanjing Shi$^1$} 
   \institute{
$^{1}$Shanghai Astronomical Observatory, Chinese Academy of Sciences, 80 Nandan Road, Shanghai 200030, China\\
$^{2}$Instituto de Física, Universidade Federal do Rio Grande do Sul, Av. Bento Gonçalves 9500, Porto Alegre, R.S. 90040-060, Brazil\\}          

   \date{Received; accepted}
   
   \titlerunning{Formation paths of massive compact ETGs}
\authorrunning{Zhu et al.}  
 
  \abstract
   {
Massive early-type galaxies (ETGs) are thought to form in two phases: an initial phase of rapid star formation and a later phase of mergers. A small fraction of these galaxies, called ``red nuggets", formed during the first phase may have survived frozen to today, having experienced no massive mergers since $z\sim 2$. Nearby massive compact ETGs are considered candidates for such relic galaxies. We study the internal dynamical structures of 15 compact ETGs with existing integral field unit (IFU) observations and 79 compact ETGs from the TNG50 simulation. We dynamically decompose each galaxy into a disk, bulge, and hot inner stellar halo, for both observations and simulations.
In TNG50, the luminosity fraction of the hot inner stellar halo (or the size of the spheroid, which includes the bulge and halo) strongly correlates with the galaxy's merger history. The true ``merger-free" galaxies show an extremely low fraction of a hot inner stellar halo (or an extremely compact spheroid). Although such compactness could result from the tidal stripping of satellites, tidal forces would also destroy the dynamically cold disk (if one exists) when the halo is removed. Thus, a galaxy is guaranteed to be merger-free if it has a very low fraction of the hot inner stellar halo and retains a dynamically cold disk.
Comparing observed galaxies with TNG50, we identify 7 of the 15 compact ETGs, PGC 11179, UGC 3816, NGC 2767, NGC 1277, PGC 32873, PGC 12562, and PGC 70520, as true merger-free galaxies. These galaxies have compact, massive bulges, likely formed through secular heating, as supported by their TNG50 analogues.
   }


   \maketitle
%

\section{Introduction}

Massive early-type galaxies (ETGs, $M_* > 10^{11}$~M$_\odot$) in the local Universe are believed to have gone through two major formation phases \citep[e.g.,][]{Naab2009, Oser2010, Du2021}. The first phase of gas collapse and star formation burst happened early (at $z\gtrsim2$) and rapidly formed a population of passively evolving objects seen at high redshift, sometimes termed ``red nuggets" \citep[e.g,][]{Daddi2005,Damjanov2009, Valentino2020}. In contrast, in the second phase, ETGs grew by accretion of material--not only through minor mergers but also through major mergers \citep{Du2021,Zhu2022a}--that built up their haloes. 
Recent observations show that a higher fraction of ETGs at redshift $z\gtrsim 3$ are disk-dominated \citep{Kartaltepe2023} and rotation-supported \citep{DEugenio2023_jwst}, whereas ETGs at the $z\sim 1$ Universe are more rotation-supported than those at $z\sim0$ \citep{Bezanson2018,DEugenio2023}. This scenario supports merger-driven growth over the past few Gyrs.
A small fraction of these red nuggets is expected to have survived ``frozen" until the present-day Universe, with no massive mergers or significant star formation since $z\sim 2$ \citep{Quilis2013ApJ...773L...8Q}. 

Massive compact ETGs are found in both fields and clusters in the local Universe \citep{Graham2015, Y17,Tortora2020,Schnorr2021, Hon2022MNRAS.514.3410H}. Some of them appear to be remarkably similar to massive quiescent galaxies at high $z$, extremely compact, and with old stellar populations ($>10$\,Gyr) \citep{Trujillo2009ApJ...692L.118T, Ferr2017MNRAS.467.1929F,Martin2019, Grebol2023MNRAS.526.4024G, Spiniello2024MNRAS.527.8793S}, thus being taken as candidates red nugget ``relics'' from the high $z$ Universe.

Studying the relic galaxies is a unique way of gaining insight into the early Universe, as they provide an ideal laboratory, frozen in time, into early galaxies. Moreover, they serve as probes for detailed measurements that are difficult to do directly at high $z$ galaxies. An interesting case is that the relic galaxy NGC 1277 \citep{Trujillo2014} has extremely bottom-heavy stellar initial mass functions (IMF) \citep{Martin2015}, a single population of metal-rich globular clusters (GC) \citep{Beasley2018}, an over-massive black hole (BH) \citep{vdB2012, Walsh2015, Walsh2016, Walsh2017}, evidence for a low dark matter (DM) fraction \citep{Comeron2023}, and has a luminous X-ray halo \citep{Fabian2013}. 
However, there are some diversities in the other candidates \citep{Y17, Ferr2017MNRAS.467.1929F}. As measured from the large sample of relic candidates in the INvestigating Stellar Population In RElics (INSPIRE) project \citep{Spiniello2024MNRAS.527.8793S}, the IMF slope is likely to be correlated with the degree of relicness \citep{Maksymowicz2024MNRAS.531.2864M}, ranked in terms of their morphological and stellar population characteristics. 
The GC population studied in 12 nearby compact ETGs has a generally higher fraction of red GCs, as expected to be relic galaxies, but still with a relatively large scatter \citep{Kang2021, Alamo-Martinez21}. BH may grow rapidly in the early Universe, and simulations predict that relic galaxies may have over-massive BHs \citep{Barber2016MNRAS.460.1147B, vanSon2019}. The BH masses have been measured in a few compact ETGs, the BH mass is indeed overmassive in NGC 1271, NGC 1270, and MRK1216 \citep{Walsh2017, Ferre2015ApJ...808...79F}, but is normal in UGC 2698 following the general black hole mass-bulge mass relation \citep{Cohn2021}.

Extremely compact and massive ETGs with uniformly old stellar populations are taken as evidence that they are relic galaxies \citep{Ferr2015, Ferr2017MNRAS.467.1929F, Martin2019}. However, a galaxy being strongly tidally stripped by the environment could also result in remnants that behave similarly in many of the aforementioned aspects; they could also be extremely compact, with uniformly old stellar populations, and present an overmassive BH \citep{vanSon2019, Barber2016MNRAS.460.1147B}. 
It is thus critically important to find the true relic galaxies that have had extreme quiescent merger histories, to distinguish them from those compact ETGs formed from strong tidal stripping. To achieve this, we need extra information to probe the merger history of these galaxies.  

However, galactic dynamics has proven to help uncover fossil records of a galaxy's merger history.
Modern cosmological simulations, such as IllustrisTNG\footnote{\url{www.tng-project.org}} and EAGLE\footnote{\url{http://eagle.strw.leidenuniv.nl/}}, have successfully reproduced a large number of galaxies with well-resolved structures that statistically match observations in many aspects, including the mass-size relation \citep{Genel2018,Lange2016,deGraaff2021},  galaxy concentration and bulge strength \citep{Rodriguez-Gomez2019}, disk scale height \citep{Zanisi2021}, and internal kinematic structures \citep{Xu2019, Zhang2025arXiv250107151Z}. The mass of a hot inner stellar halo, dynamically defined by the stellar orbits with circularity $\lambda_z < 0.5$ and at $r>3.5$ kpc, is found to be highly correlated with the galaxies' accreted stellar mass, especially for galaxies having massive major mergers \citep{Zhu2022a, Zhu2022b}, and the correlation is almost identical in TNG100, TNG50, and EAGLE, independent of resolution and galaxy formation model. However, there is relatively large scatter in the compact ETGs with a low fraction of hot inner stellar halo, and detailed studies are needed to further understand such scatter.

The spatial resolutions of TNG100 and EAGLE are not high enough to study the internal structures of these extremely compact galaxies. However, TNG50 has the highest resolution and still has a large number of galaxies in different environments \citep{Nelson2019release, Pillepich2019}. In fact, relic-like galaxies are found in TNG50 \citep{Flores2022, Deeley2023, Moura2024} in different environments. In this paper, we study the internal dynamical structures of 15 compact ETGs from observations \citep{Y17} and a comparable sample of compact ETGs from TNG50, defining their dynamical structures exactly the same way.
By understanding the true assembly histories of these compact ETGs in TNG50, we propose key structural properties that can distinguish true "merger-free" galaxies from other compact ETGs. Combining stellar populations and our analysis on dynamical
structures will provide decisive criteria for identifying clean relic galaxies with extremely quiescent merger histories from observational data.

The paper is organised as follows. In Section 2, we describe the sample of compact ETGs from observations and simulations and define their internal dynamical structures. In Section 3, we directly compare the internal structure of observed and simulated compact ETGs with those from TNG50, using their known assembly histories to identify key evidence for distinguishing true merger-free galaxies. In Section 4, we illustrate two different formation pathways of compact ETGs in TNG50 by examining two typical cases. We discuss our findings in Section 5 and present our conclusions in Section 6.

\begin{figure*}
\centering\includegraphics[width=18cm]{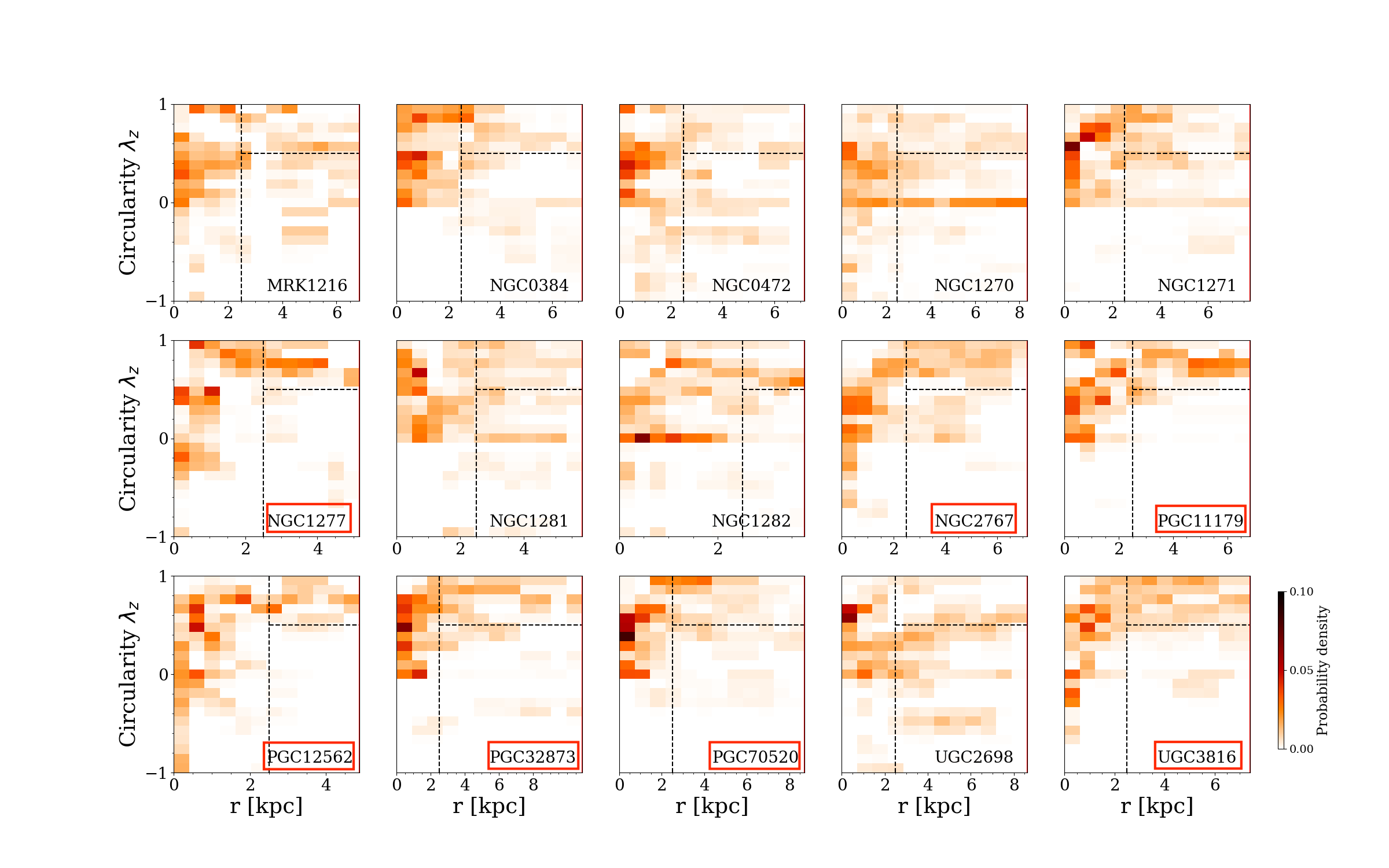}
\caption{Stellar orbit distribution of 15 compact ETGs \citep{Y17} obtained from the best-fitting orbit superposition model. The colour represents the probability density of the stellar orbits in the phase space of the time-averaged radius $r$ versus circularity $\lambda_z$, $p(r, \lambda_z)$, as indicated by the colour bar. The vertical dashed lines indicate $r=2.5$ kpc, while the horizontal line indicates $\lambda_z = 0.5$. We define the bulge, disk, and the hot inner stellar halo within $r_{\rm max} = 7$ kpc, as illustrated in the top left panel. The seven galaxies marked in red squares have a significant fraction of disk and a very low fraction of hot inner stellar halo.
}
\label{fig:rlz15}
\end{figure*}
\section{Compact ETGs from observations and simulations}
\label{s:obs}
We study compact ETGs from observations and from the cosmological simulation TNG50.  
In this section, we will introduce the general properties of the galaxy sample and the dynamical structure decomposition, first for observed galaxies and then for simulated galaxies. We take the same approach of describing the dynamical structures of observed and simulated galaxies, thus making them directly comparable. Note that the dynamical decomposition is based on the stellar orbit distribution, which is obtained through different ways for observed and simulated galaxies, and we will introduce them separately.

\subsection{Compact ETGs from observations}

\subsubsection{Sample from observations}

We take 15 well-studied massive compact ETGs from observations \citep{Y17}. These galaxies are observed by HST and the wide-filed IFU instrument PPAK \citep{Verheijen2004AN....325..151V, Kelz2006PASP..118..129K}, they are proved to be massive, compact, and with uniformly old stellar populations, while they have a variety of stellar kinematic features. 

\begin{table*}
\caption{Basic information of the 15 massive compact ETGs.}
\scriptsize\centering
\label{tab:15compact}
\begin{tabular}{*{12}{l}}
\hline
       Galaxy name  &  $\log(M_{*})$ & $R_e$ [kpc] & $d$ [Mpc] & $R_{\rm max}$ [kpc] & $f_{*,\rm disk}$ &  $f_{*,\rm halo}$ & $f_{*,\rm bulge}$ & $\log(M_{*,\rm Sph})$ & $R_{\rm Sph}$ [kpc]  & Environment & merger-free? \\
\hline
MRK1216 & 11.34 &   3.0 &    94 &   6.8 &  0.19 &  0.21 &  0.60 & 11.22 & 1.9 & isolated & Maybe\\

NGC0384 & 10.96 &   1.8 &    59 &   7.2 &  0.17 &  0.14 &  0.69 & 10.69 & 0.9 &group & Maybe\\ 

NGC0472 & 11.07 &   2.4 &    74 &   7.2 &  0.11 &  0.21 &  0.68 & 10.98 & 1.5 &group/isolated & Maybe\\

NGC1270 & 11.31 &   2.2 &    69 &   8.4 &  0.12 &  0.31 &  0.57 & 11.17 & 1.5 & group & No\\

NGC1271 & 11.06 &   2.0 &    80 &   7.8 &  0.21 &  0.19 &  0.60 & 10.64 & 1.3 & group & Maybe\\

NGC1277 & 11.13 &   1.3 &    71 &   5.2 &  0.25 &  0.05 &  0.70 & 10.77 & 0.54 & group & Yes\\

NGC1281 & 11.00 &   1.6 &    60 &   5.8 &  0.16 &  0.22 &  0.62 & 10.76 & 1.2 &group & Maybe \\

NGC1282 & 10.77 &   1.3 &    31 &   3.8 &  0.11 &  0.07 &  0.82 & 10.58 & 0.64 & group & Not sure\\

NGC2767 & 11.12 &   2.4 &    74 &   7.2 &  0.31 &  0.14 &  0.55 & 10.85 & 0.76 & group & Yes\\

PGC11179 & 11.16 &   2.1 &    94 &   6.8 &  0.35 &  0.05 &  0.60 & 10.75& 0.71 &group & Yes\\

PGC12562 & 10.74 &   0.7 &    67 &   4.9 &  0.17 &  0.05 &  0.78 & 10.46 & 0.51&group & Yes\\

PGC32873 & 11.28 &   2.3 &   112 &  10.9 &  0.26 &  0.09 &  0.65 & 10.90 & 1.2 &isolated & Yes\\

PGC70520 & 10.95 &   1.6 &    72 &   8.7 &  0.23 &  0.11 &  0.66 & 10.67 & 0.74 & isolated & Yes\\

UGC2698 & 11.58 &   3.7 &    89 &   8.6 &  0.11 &  0.38 &  0.51 & 11.41& 2.38 &group & No \\

UGC3816 & 10.96 &   2.1 &    51 &   7.4 &  0.33 &  0.10 &  0.56 & 10.60 & 0.77 &group/isolated & Yes\\
\hline
\hline
 \end{tabular}
 \tablefoot{The columns from left to right are the galaxy name, stellar mass $\log(M_{*}/M_{\odot})$, half-light radius $R_e$ in [kpc], distance $d$ in Mpc, the spatial coverage of the IFU data $R_{\rm max}$ in kpc, the luminosity fraction of disk $f_{*,\rm disk}$, hot inner stellar halo $f_{*,\rm halo}$, and bulge $f_{*,\rm bulge}$, stellar mass of the hot spheroid $\log(M_{*,\rm Sph}/M_{\odot})$, half-light radius of the hot spheroid $R_{\rm Sph}$ in kpc, environments, and merger histories of these galaxies according to our analysis of their internal orbit distribution (Section~\ref{S:distinguish}).}
\end{table*}

The 15 galaxies from \citet{Y17} were observed by the wide-field IFU instrument PPAK \citep{Verheijen2004AN....325..151V, Kelz2006PASP..118..129K} on the 3.5-m telescope at Calar Alto. The luminosity-weighted stellar kinematic maps, including velocity $v$ and velocity dispersion $\sigma$, are derived by the pPXF \citep{Cappellari2017} fitting to the spectra at a wavelength range of 4200-7000\AA \citep{Y17}. These galaxies were also observed by HST, which provides a high-resolution photometry image. Here we use exactly the same kinematic maps and photometric data as published in \cite{Y17}.

\subsubsection{Orbit-superposition models}

We obtain the stellar orbit distribution of each galaxy by creating Schwarzschild's orbit-superposition models, which simultaneously fits the luminosity density distribution and the kinematic maps of the galaxies. Details of the method can be found in \cite{vdB2008, Zhu2018a}. The method has been further extended to include stellar age and metallicity \citep{Zhu2020, Ding2023A&A...672A..84D, Jin2024A&A...681A..95J}, and to model barred galaxies \citep{Tahmasebzadeh2022ApJ...941..109T, Tahmasebzadeh2024MNRAS.534..861T}. The orbit-superposition models for these 15 compact ETGs have already been constructed in \citet{Y17}, here we have exactly the same data and generally the same model construction. We briefly describe some main steps in the following paragraphs.

The major steps of constructing the orbit superposition model include:(1) construction of a model for the gravitational potential with a few free parameters, (2) calculation of an orbit library in the gravitational potential, and (3) inferring the weights of orbits by fitting the model to the data including the luminosity density and kinematic maps. We calculate hundreds of models by exploring the free parameters in the gravitational potential and find the best-fitting models with minimum $\chi^2$ between the model and the data.
In the end, we obtain 3D models superposed by stellar orbits that match all the data well. 

The gravitational potential is generated by the combination of stellar mass, dark matter mass, and a central black hole fixed at $10^{8}$\,\Msun\,. Although the black hole remains unresolved by the kinematic data and does not affect our results. 
To obtain the stellar mass distribution, we first fit the HST $H$-band image with a Multiple Gaussian Expansion (MGE) model \citep{Cappellari2002}. 
Then by adopting a set of viewing angles ($\vartheta, \psi, \phi$), we deprojected toward a triaxial MGE model which represents the intrinsic stellar luminosity density. 
Finally, by multiplying a constant stellar mass-to-light ratio $M/L$, we arrive at the intrinsic stellar mass distribution.

The three viewing angles are directly related to the three axis ratios $(p,q,u)$ that describe the intrinsic 3D shape of the stellar luminosity distribution \citep{vdB2008}.
We leave the intrinsic intermediate-to-major and minor-to-major axis ratios $(p,q)$ as free parameters but limit $u=0.98-0.9999$, so moderate triaxiality is allowed in the model.
The dark matter halo is assumed to be spherical with an NFW \citep{nfw1997} radial profile with two free parameters, the halo mass $M_{200}$ defined as the total mass enclosed within the Virial radius $r_{200}$ within which the average density is 200 times the critical density, and the concentration $c$ defined as the ratio between the dark matter Virial radius and the dark matter scale radius. The model construction is generally the same as that adopted in \citet{Y17}.
In total, we have five free hyperparameters in the model: $M/L$, $p$, $q$, $M_{200}$, and $c$.

For each set of hyperparameters, we computed tens of thousands of orbits in the corresponding gravitational potential. 
Orbit sampling and calculation follow what we described in \citet{Zhu2018a}. 
The weights of the orbits in the resulting orbit library were then determined by fitting simultaneously the orbit superposition model to the intrinsic and projected luminosity density and stellar kinematic maps. 

We calculate hundreds of models with different collections of hyperparameters. 
Once we have explored the hyperparameter space, we selected the best-fitting models within the $1\sigma$ confidence level defined as: 
$\Delta\chi^2 \equiv \chi^2- \min(\chi^2) < \sqrt{2 \times n_{\rm GH} \times N_{\rm obs}}$, 
where $n_\mathrm{GH}=4$ is the number of stellar kinematic moments and $N_\mathrm{obs}$ is the number of bins of each kinematic map.  
There are usually a few tens of models with different values of the hyperparameters within this $1\sigma$ confidence level for each galaxy, and all the models within the $1\sigma$ confidence fit the luminosity density and kinematic maps reasonably well.  
In this paper, we only take the best-fitting model with minimum $\chi^2$ for the analysis. 

Our method yields the weights of the different stellar orbits that contribute to the best-fitting model. We then describe the stellar orbit distribution of the model as the probability density of the stellar orbit in the phase space of $r$ versus circularity $\lambda_z$ following \citet{Zhu2018a}.
We characterise each stellar orbit with two main properties: the time-averaged radius, $r$, which represents the extent of the orbit, and the circularity, $\lambda_z = J_z/J_{\rm max} (E)$, which represents the angular momentum of the orbit around the minor $z$-axis normalised by the maximum of a circular orbit with the same binding energy $E$.
Whereas $|\lambda_z| \sim 1$ represents dynamically cold orbits dominated by regular rotation, $\lambda_z \sim 0$ represents dynamically hot orbits dominated by radial random motions or long-axis tubes with regular rotation around the long axis.
Negative values $\lambda_z<0$ refer to orbits that counter-rotate with respect to the net (prograde) rotation. 
Figure~\ref{fig:rlz15} presents the resulting stellar orbit distributions of the 15 compact ETGs we obtained, as probability density distributions $p(r, \lambda_z)$ in the phase space of $r$ versus $\lambda_z$ for each galaxy. Although we used HST H band images to constrain the surface brightness, the internal stellar orbit distribution is mainly determined by the stellar kinematic data, which were taken in 4200-7000\AA\, and luminosity-weighted, and we take the r band as the closest approximation. Our dynamical model and the resulting stellar orbit distribution are thus also luminosity-weighted in the r-band.

\begin{figure}
\centering\includegraphics[width=8cm]{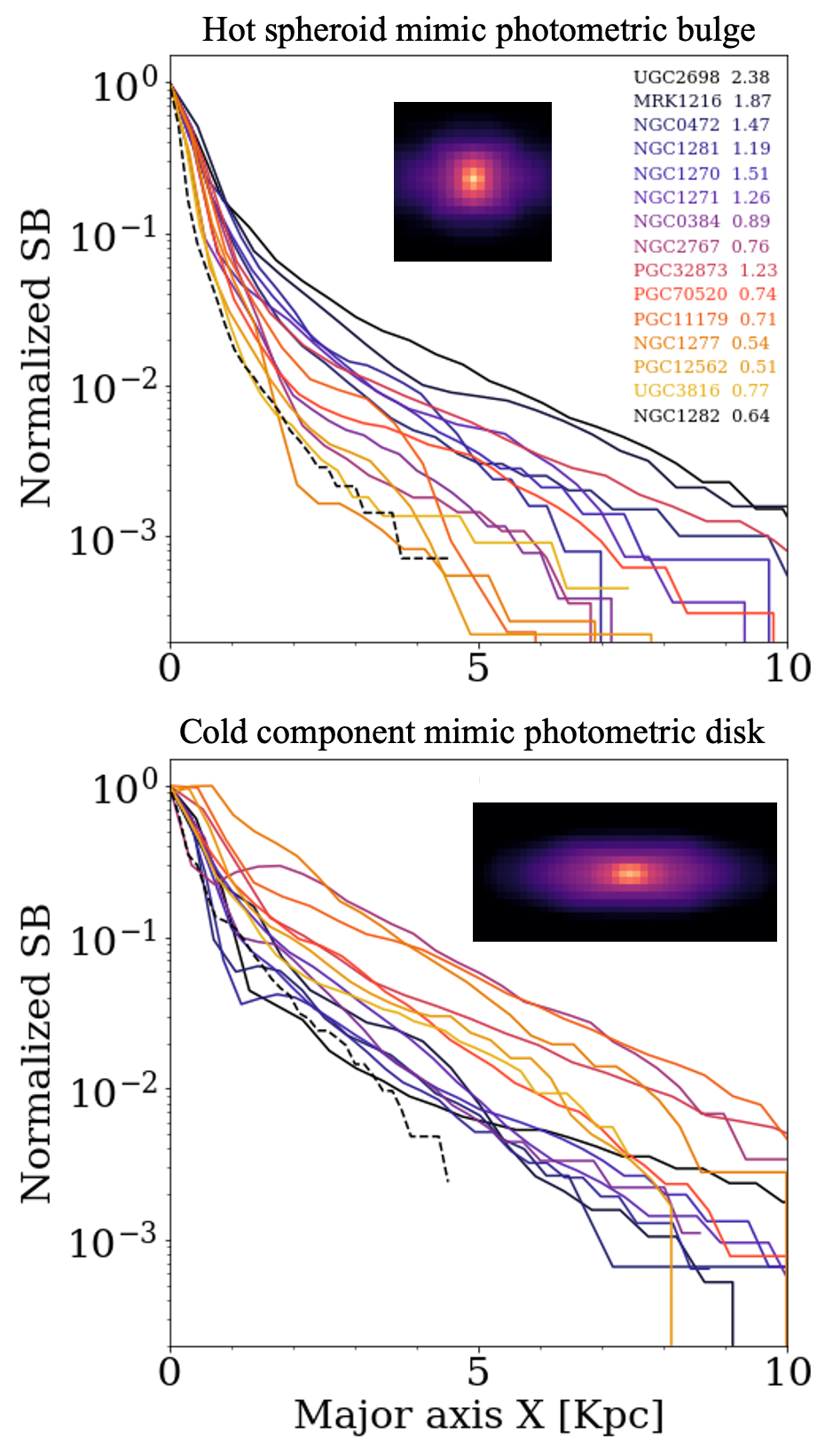}
\caption{The surface brightness profile of the dynamically hot spheroid (defined by all orbits with $\lambda_z<0.5$) and dynamically cold component (defined by all orbits with $\lambda_z>0.5$) of the 15 compact ETGs. Galaxy names and half-light radius (in kpc) of their hot spheroids are labeled in the top panel, the dash curve represents NGC 1282 with limited data coverage. The inset panel represents the 2D surface brightness of the hot spheroid of one typical galaxy NGC 1277. 
}
\label{fig:SB15}
\end{figure}

\subsubsection{Dynamical decomposition and quantification of the hot component}
\label{ss:obs_decomp}

We take two approaches to quantify the dynamically hot component of the galaxy. The first approach involves the decomposition of each galaxy into three stellar components based on their stellar orbit distribution: 
\begin{itemize}
\item disk ($\lambda_z>0.5$, $\rcut<r<\rmax$), 
\item hot inner stellar halo ($\lambda_z < 0.5$, $\rcut<r<\rmax$),
\item bulge ($r <\rcut$).
\end{itemize}
Here we choose $\rcut = 2.5$ kpc empirically from Fig.~\ref{fig:rlz15} and the transition radius from bulge to halo in TNG50 for these compact galaxies, we choose $\rmax = \max(7 {\rm kpc},  2 R_e)$, since most of the 15 galaxies have data coverage extending to this radius. 
We note that the definition of bulge, disk, and hot inner stellar halo is decided considering the special properties of compact ETGs, and is slightly different from that for the general population of all TNG50 galaxies in \citet{Zhu2022a}. This kind of decomposition is based on the internal stellar orbit distribution and thus can only be done for real galaxies with IFU observations associated with dynamical models. 

To make it more comparable to the results of the morphological decomposition based on photometric images, we take a second approach by considering all orbits with $\lambda_z<0.5$ as a dynamically hot spheroid and $\lambda_z>0.5$ as a cold component. These are comparable with the bulge/spheroid and disk from photometric bulge-disk decomposition \citep{Zhu2018c}. We rebuilt the surface density distribution of the hot spheroid and cold component by sampling particles from the orbits. The surface brightness profiles of the two components along the major axis of the 15 galaxies are shown in Figure~\ref{fig:SB15}. In this sample, the galaxies with the most compact hot spheroids are associated with spatially extended cold components.
We calculate the half-light radius of the hot spheroid component from the 2D surface brightness map, the mass and size of the hot spheroid are included in table~\ref{tab:15compact}. 

\begin{figure}
\centering\includegraphics[width=8cm]{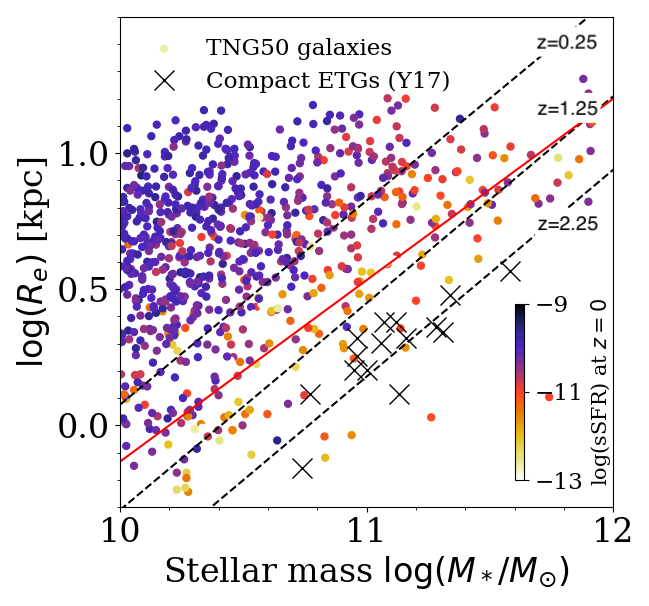}
\caption{Compact ETGs from observations and from Illustris TNG50 in the plane of stellar mass $M_*$ versus size $R_{\rm e}$. The dots are TNG50 galaxies coloured by their specific star formation rate $\log({\rm sSFR})$ at $z=0$.
Grey `x' symbols are the 15 compact ETGs from observations \citep{Y17}. The black dashed lines indicate the mass-size relations for galaxies observed at $z=0.25$, $z=1.25$, and $z=2.25$ \citep{vdw2014}. We select compact ETGs from TNG50 below the red line with $M_*/ R_e^{1.5} \equiv \Sigma_{1.5} > 10^{10.2}M_{\odot}{\rm kpc}^{-1.5}$ \citep{Barro2013} and with $\log({\rm sSFR}) < -11$.
}
\label{fig:MRe_fig1}
\end{figure}

\subsection{Compact ETGs in TNG50}
\label{s:tng50}

\subsubsection{Sample selection from TNG50}
We use the cosmological simulation TNG50, which has a stellar mass resolution of $8.5\times 10^4$ \,\Msun\, softening length of 0.3 kpc and a relatively large cosmological volume of 50 Mpc.
For a simulated galaxy, we take the total stellar mass within 30 kpc as the stellar mass $M_*$. To measure the half-light radius mimic observations, we randomly project each galaxy onto the 2D observational plane, create an r-band-weighted surface brightness map, and measure the half-light radius $R_e$ from the 2D map. All TNG50 galaxies with $M_* > 10^{10}\,M_{\odot}$ are shown in Fig.1.
We select compact ETGs from TNG50 with $M_* > 2\times 10^{10}\,M_{\odot}$, $R_e<R_{\rm crit}$, and with a specific star formation rate $\log({\rm sSFR}) < -11$ at $z = 0$.
We define compact galaxies with $M_*/ R_e^{1.5} \equiv \Sigma_{1.5} > 10^{10.2}M_{\odot}{\rm kpc}^{-1.5}$, which generally follows the definition in \citet{Barro2013}, but we choose a less strict criterion of 10.2 rather than 10.3 adopted in \citet{Barro2013} to enlarge the sample size.

This allows us to have a sample of 90 galaxies which is representative for galaxies in different environments and with different internal structures. Many of them have a smaller stellar mass and a slightly larger size than the 15 compact ETGs from observations \citep{Y17}.
There are various definitions of "compactness" throughout the literature, each considering more or less restrictive criteria (see Table 3 of \citet{Lisiecki2023A&A...669A..95L}). The sample adopted here still defines the galaxy as compact, while ensuring a sufficiently large number of galaxies in TNG50 to allow meaningful comparisons with observations.

We further exclude three galaxies with ongoing mergers or with recent star formation blobs telling from their mock photometric images.
To further remove galaxies with significantly young stars, we calculate the luminosity-weighted average stellar age map across the 2D plane, and we require the stellar age to be larger than $5$ Gyr at the outer regions ($r>2$ kpc) where disks dominate. It excludes another eight galaxies. Here we adopted a relatively loose cut of $>5$ Gyr to maintain a generous sample. In total, we have 79 compact ETGs from TNG50. We define the most massive galaxy in a halo as the central galaxy, and the rest as satellite galaxies. The 79 compact ETGs are located in different environments, 35 of them are central galaxies, and 44 of them are satellites.

\begin{figure*}
\centering\includegraphics[width=18cm]{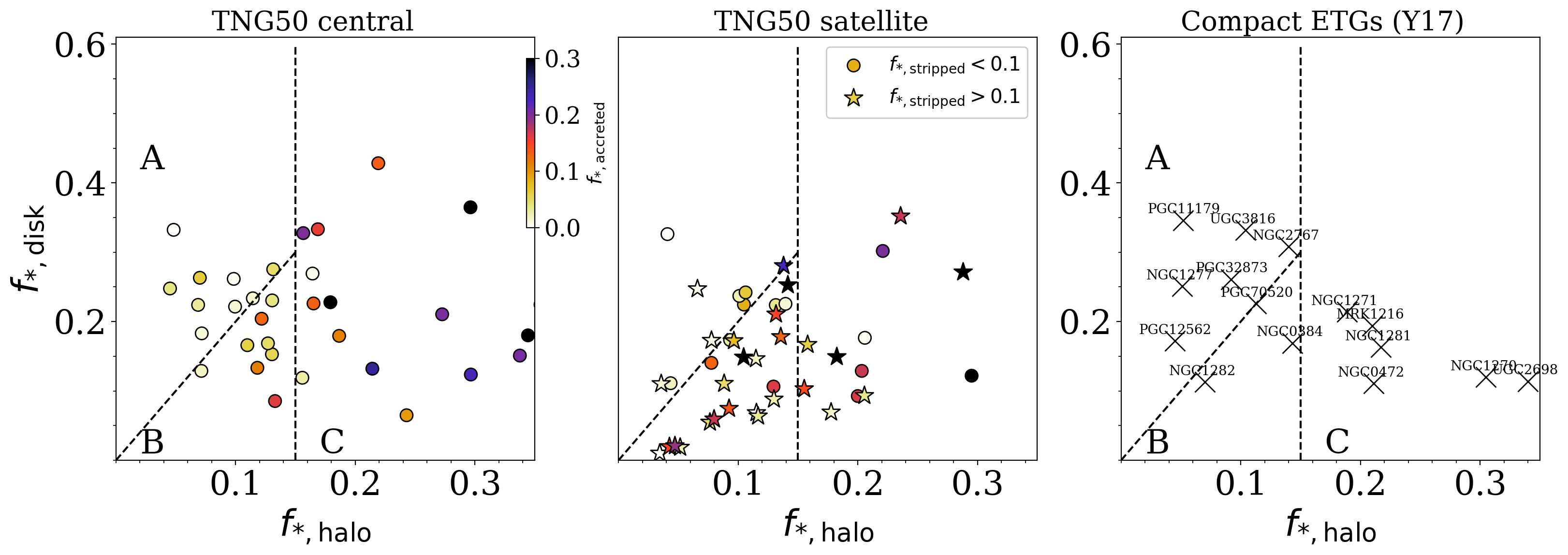}
\caption{The r-band luminosity fraction of $f_{*,\rm halo}$ versus $f_{*, \rm disk}$. 
{\bf Left:} Compact ETGs selected from TNG50 central galaxies, coloured by the accreted stellar mass fraction $f_{*,\rm accreted}$ as indicated by the corlorbar, all these galaxies are not strongly stripped. {\bf Middle:} compact ETGs selected from TNG50 satellite galaxies, with $f_{*,\rm accreted}$ coloured by the same color scheme. The dots and stars indicate the galaxies with stripped stellar mass fraction $f_{*,\rm stripped} <0.1$ and $f_{*,\rm stripped} >0.1$, respectively. We define regions A, B, and C to guide our comparison between simulations and observations. All the TNG50 compact ETGs located in region A have extremely quiescent merger histories $f_{*,\rm accreted}<0.1$, for both central and satellite galaxies.
{\bf Right:} The 15 compact ETGs from observations. Compared with TNG50 analogues, the seven galaxies located in region A are likely to have extreme quiescent merger histories.
}
\label{fig:ff}
\end{figure*}

\subsubsection{Assembly history}

The assembly histories of all galaxies in the simulations can be followed via the merger trees, as constructed by the Sublinkgal code based on the baryonic component of the subhalos \citep{Rodriguez2015}. In this algorithm, each galaxy is assigned a unique descendant. 

For each compact ETG in our sample, we define two key parameters from the merger tree: the accreted stellar mass fraction $f_{*,\rm accreted}$ and the stripped stellar mass fraction $f_{*, \rm stripped}$. We define the accreted fraction as the ratio of the total stellar mass of all accreted satellite galaxies to the stellar mass of the main progenitor galaxy at $z=0$:
\begin{equation}
f_{*,\rm accreted} = (M_{*,\rm ex1} + M_{*,\rm ex2} +...)/M_{*, z=0},
\end{equation}
 where $M_{*,\rm ex1}$, $M_{*,\rm ex2}$,... are the mass of all the accreted satellite galaxies identified from the merger tree. 
We define the stripped stellar mass fraction as 
\begin{equation}
    f_{*, \rm stripped} = {M_{*,\rm max}}/M_{*,\rm z=0} - 1,    
\end{equation}
where $M_{*,\rm z=0}$ is the galaxy's stellar mass at $z=0$, and ${M_{*,\rm max}}$ is the maximum stellar mass of its main progenitor that has even been reached. 

Note that the accreted stellar mass fractions roughly include the mass of all stars that have ever been accreted into the galaxy. In a galaxy that is not significantly stripped, $f_{*,\rm accreted}$ generally equals the ex-situ fraction of stars in the galaxy. For a strongly stripped galaxy, it might have experienced massive mergers, but most of the accreted stars were then stripped with the outer envelope. In this case, the ex-situ fraction of stars in the current galaxy might be very low, but the accreted fraction $f_{*,\rm accreted}$ defined from the merger tree will be high, reflecting the merger events that the galaxy has ever experienced. 

The 79 compact ETGs from TNG50 have a large variation in assembly histories as discussed in \cite{Moura2024}: some of them indeed have extremely quiescent merger histories with low $f_{*,\rm accreted}$; some have actually experienced significant mergers with relatively high $f_{*,\rm accreted}$; all of the compact ETGs selected from TNG50 central galaxies are not strongly stripped with $f_{*, \rm stripped}<0.1$, while a large fraction of these compact ETGs selected from satellites in cluster environments were heavily stripped with high $f_{*, \rm stripped}$.

\subsubsection{Dynamical decomposition of simulated galaxies}

For the simulated galaxies, we have the 6D phase-space information of the stellar particles instantaneously at $z=0$. We adopt an approximate method to obtain the time-averaged $r$ and $\lambda_z$ along the orbits, whereby the averaging is done in phase space. In practise, we assume stellar particles with similar energy $E$ and angular momentum $L_z$ to be on similar orbits: hence we measure the average $r$ and $\lambda_z$ across such particles from the simulation and adopt them as the orbital $r$ and $\lambda_z$ values. 

We take this approach, as it is fast and approved to produce consistent results in several previous studies \citep{Zhu2022a,Zhu2018b}. 
The stellar orbit distribution of a typical simulated galaxy as a probability density distribution of particles in $r$ versus $\lambda_z$ can then be obtained, we weight the stellar particles by the r-band luminosity, in alignment with the observations.

We then take the same approaches to characterise the hot component: (1) define the bulge, disk, and hot inner stellar halo and calculate the r-band luminosity fraction of each component; (2) define the hot spheroid component and calculate its mass and size from the projected r-band surface brightness map, both in exactly the same way as for the real galaxies.

\begin{figure*}
\centering\includegraphics[width=12cm]{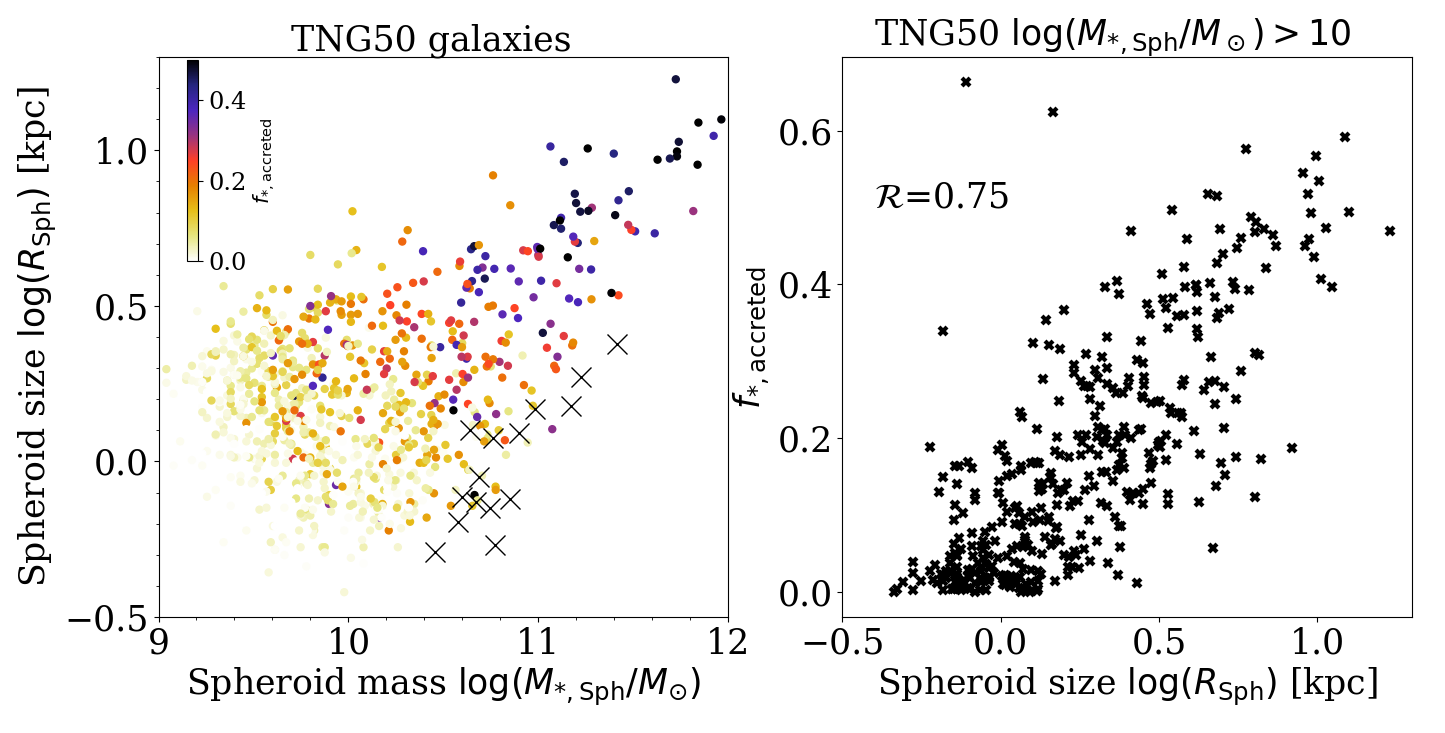}
\caption{Size of the dynamically hot spheroid as an indicator of accreted stellar mass fraction $f_{*,\rm accreted}$. {\bf Left:} Spheroid mass $M_{*,\rm Sph}$ versus spheroid size $R_{\rm Sph}$, the dots are TNG50 galaxies coloured by their accreted fraction $f_{*,\rm accreted}$. The black symbols 'x' indicate the 15 compact ETGs from observations \citep{Y17} as listed in Table~\ref{tab:15compact}. {\bf Right:} Spheroid size $R_{\rm Sph}$ versus accreted fraction $f_{*,\rm accreted}$ for TNG50 galaxies with $M_{*,\rm Sph} > 10^{10}\,$\Msun. }
\label{fig:MRe_sph}
\end{figure*}

\begin{figure*}
\centering\includegraphics[width=18cm]{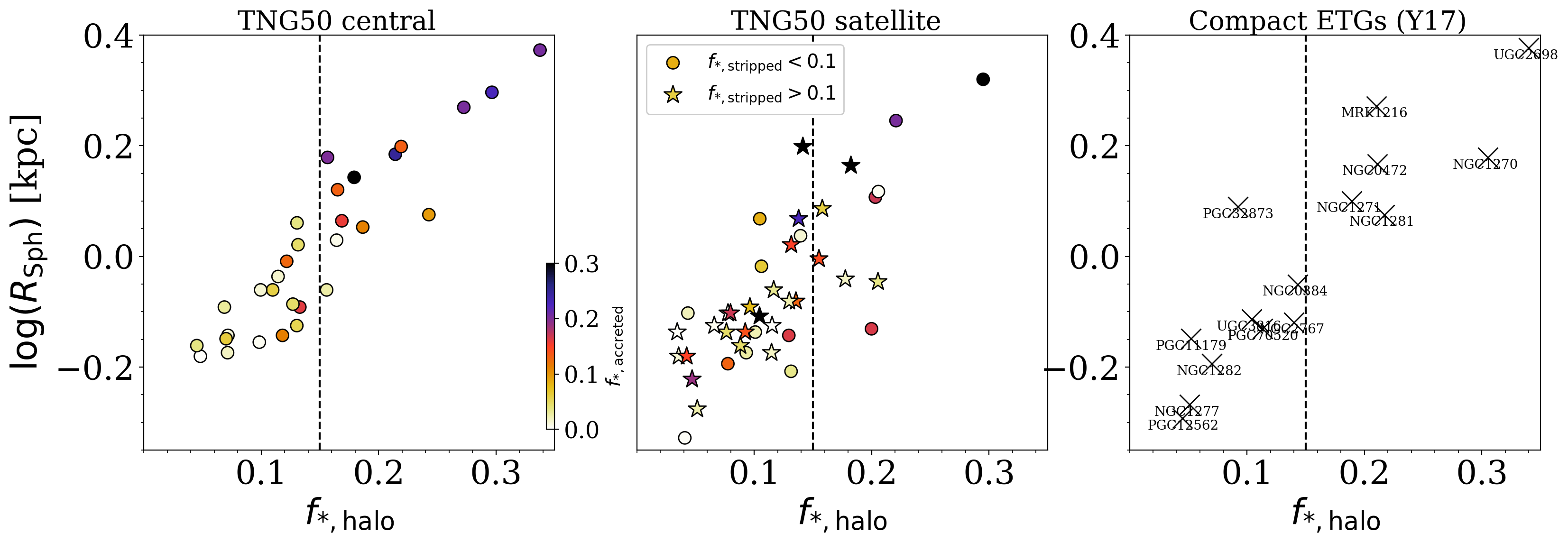}
\caption{The luminosity fraction of hot inner stellar halo $f_{*,\rm halo}$ versus the spheroid size $R_{\rm Sph}$. {\bf Left:} Compact ETGs selected from TNG50 central galaxies, coloured by the accreted stellar mass fraction $f_{*,\rm accreted}$ as indicated by the corlorbar. {\bf Middle:} compact ETGs selected from TNG50 satellite galaxies, with $f_{*,\rm accreted}$ coloured by the same color scheme. {\bf Right:} The 15 compact ETGs from observations. The vertical dashed line indicates $f_{*,\rm halo} = 0.15$. The spheroid size and $f_{*,\rm halo}$ are strongly correlated for both simulated and observed galaxies. The two parameters work similarly well for indicating the $f_{*,\rm accreted}$ of central galaxies.}
\label{fig:Rsphfsat}
\end{figure*}

\section{Distinguishing `merger-free' compact ETGs from stripped galaxies}
\label{S:distinguish}
With the internal dynamical structures well described in exactly the same way for observed and simulated galaxies, we try to find some key structural properties probing the merger history of compact ETGs using the simulations. We gain insights into the merger history of observed compact ETGs by comparing them with their simulation analogues.

\subsection{The luminosity fraction of disk versus. hot inner stellar halo}
\label{ss:fdiskfhalo}

In Fig.~\ref{fig:ff}, we show the luminosity fraction of $f_{*,\rm disk}$ versus $f_{*,\rm halo}$, the three panels from left to right are the compact ETGs selected from the TNG50 central galaxies, compact ETGs from the TNG50 satellite galaxies, and the 15 compact ETGs from observations. 

To guide our comparison between simulations and observations, we define regions A, B, and C based on the distribution of TNG50 galaxies with different accreted stellar mass fractions $f_{*,\rm accreted}$ in this figure: Regions A and B have $f_{*,\rm halo}<0.15$ and are further separated by the line with $f_{*,\rm disk}= 2f_{*,\rm halo}$, while region C has $f_{*,\rm halo}>0.15$.

For compact ETGs selected from TNG50 central galaxies, the accreted stellar mass fraction of galaxies $f_{*,\rm accreted}$ is highly correlated with the luminosity fraction of the hot inner stellar halo $f_{*,\rm halo}$. 
All galaxies located in region A have extremely quiescent merger histories with $f_{*,\rm accreted} < 0.1$, and most galaxies in region C have relatively higher accreted stellar mass fractions; there are not many galaxies in region B, and most of them also have low $f_{*,\rm accreted}$.

For the compact ETGs selected from the TNG50 satellites, the number of galaxies in regions A and C is significantly reduced. A significant fraction of galaxies are strongly stripped and located in region B, with stars both in their disk and hot inner stellar halo stripped. Galaxies that remain in region A are still those with quiescent merger histories ($f_{*,\rm accreted} < 0.1$), including those partially stripped or not stripped. Galaxies in regions B and C are a mixture of galaxies with different merger histories, with their internal structures stripped or heated by the cluster environments.

In summary, the compact ETGs are guaranteed to be truly merger-free if they are located in region A, for both central and satellite galaxies. Central galaxies are still highly likely to be merger-free in region B, and are unlikely to be merger-free in region C. It is hard to tell their assembly history for satellite galaxies if they are located in region B or C. The different locations of galaxies in this figure with different merger histories are consistent with that found in \citet{Moura2024}.

For the 15 compact ETGs from observations \citep{Y17}, most of them are in group environments (see \citealt{Alamo-Martinez21}), some might be identified as satellite galaxies align with the definition from cosmological simulations; however, they are much more massive than the satellite galaxies selected from TNG50, and there is no observational evidence for them to be strongly stripped. These 15 compact ETGs are mostly located in regions A and C, similar to the sample of central galaxies selected from TNG50. 
We have seven of them, PGC 11179, UGC 3816, NGC 2767, NGC 1277, PGC32873, PGC 12562, PGC70520, located in region A, thus highly likely to be true merger-free with a merger fraction $f_{*,\rm accreted} < 0.1$. 
The rest of the galaxies are likely to have experienced a relatively higher fraction of mergers that altered their internal structures.

\subsection{The size of dynamically hot spheroid}
To generally show the correlation between the size of a dynamically hot spheroid and the galaxy's merger histories, we first take all TNG50 galaxies with $M_{*}>10^{10}$\,\Msun, define the dynamically hot spheroid by stars with $\lambda_z < 0.5$ for each galaxy, and calculate the mass $M_{*,\rm Sph}$ and the half-light radius $R_{\rm Sph}$ of the hot spheroid component. In Fig.~\ref{fig:MRe_sph}, we show the spheroid mass versus spheroid size for all TNG50 galaxies coloured by their accreted stellar mass fraction $f_{*,\rm accreted}$. For galaxies with $M_{*,\rm Sph} > 10^{10}$\,\Msun, their accreted stellar mass fraction $f_{*,\rm accreted}$ is highly correlated with the spheroid size, the correlation is much stronger than with galaxy size without decomposing the disk and is also stronger than with spheroid mass. 
For all galaxies with $M_{*,\rm Sph} > 10^{10}$\,\Msun, the correlation between the accreted stellar mass fraction and the spheroid size has a high Pearson correlation coefficient of $\mathcal{R}(f_{*,\rm accreted}, R_{\rm Sph}) = 0.75$.

We further check how well $R_{\rm Sph}$ can indicate the merger history of compact ETGs, as replacement for $f_{*,\rm halo}$. In Fig.~\ref{fig:Rsphfsat}, we show the correlation between the spheroid size $R_{\rm Sph}$ and the luminosity fraction of the hot inner stellar halo $f_{*,\rm halo}$, coloured by $f_{*,\rm accreted}$, for compact ETGs selected from TNG50 central galaxies, TNG50 satellite galaxies, and observations. The spheroid size $R_{\rm Sph}$ is indeed highly correlated with the luminosity fraction of the hot inner stellar halo $f_{*,\rm halo}$ for all galaxies. 

For compact ETGs selected from TNG50 central galaxies, galaxies with either low fractions of hot inner stellar halo or extremely compact spheroids have quiescent merger histories ($f_{*,\rm accreted}<0.1$). However, for satellite galaxies, those with low $f_{*,\rm halo}$ or small $R_{\rm Sph}$ are still a mix bag of galaxies with different $f_{*,\rm accreted}$; they could have frequent merger histories, but the outer envelopes have been stripped. Thus, we still need the selection of the disk fraction ($f_{*, \rm disk}$) to distinguish the true merger-free galaxies with internal structures not significantly altered by mergers or cluster environments.

The seven galaxies from observations highly likely to be true merger-free galaxies (PGC 11179, UGC 3816, NGC 2767, NGC 1277, PGC32873, PGC 12562, PGC70520 located in region A of Fig.~\ref{fig:ff}) also have extremely compact spheroids. In addition, the spheroid size of NGC 1282 is also very small, but it may have a large uncertainty due to the limited IFU data coverage of this galaxy.

\section{Illustration of different formation paths of compact ETGs in TNG50}

\begin{figure*}
\centering\includegraphics[width=16cm]{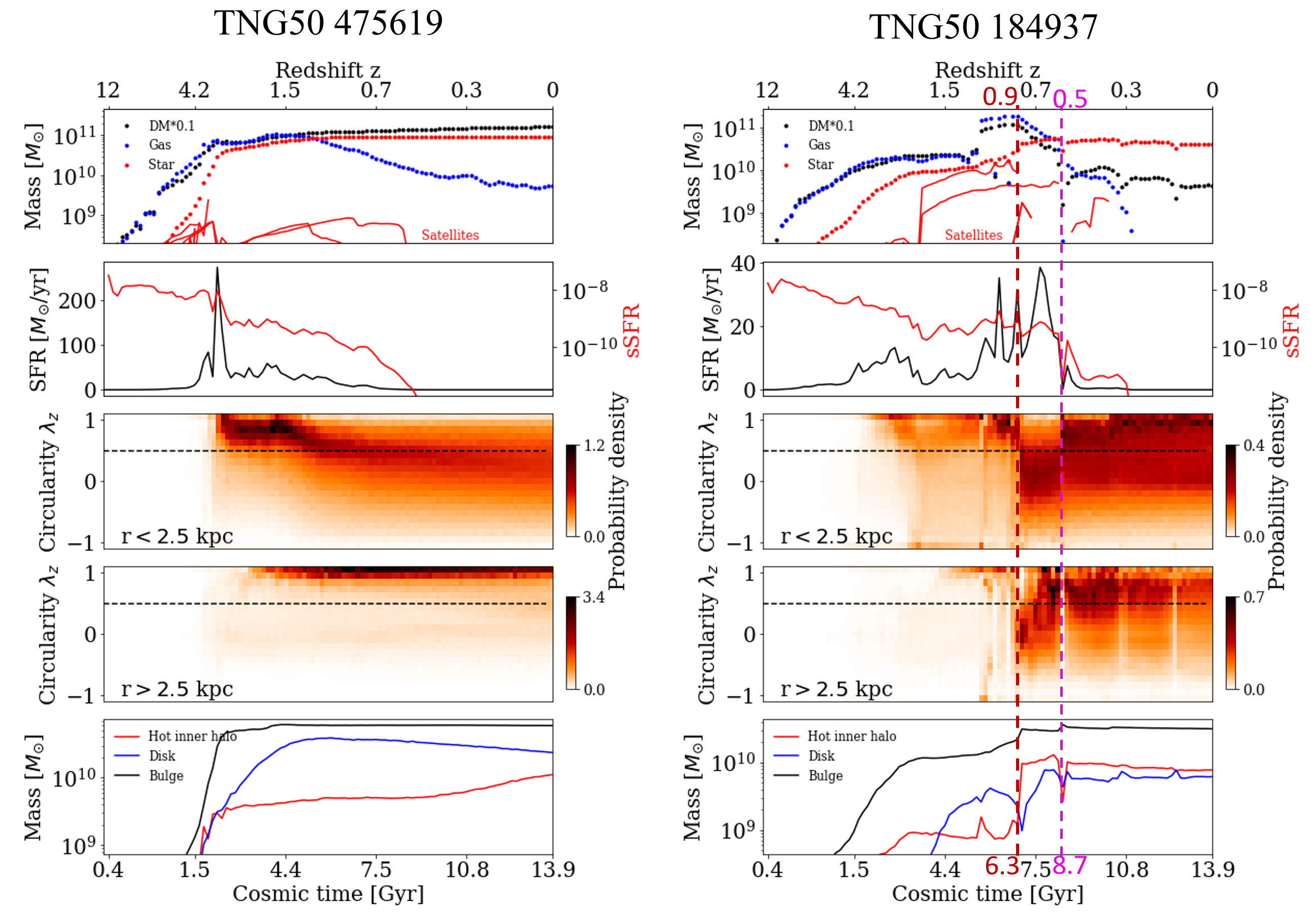}
\caption{The assembly history of two typical compact ETGs in TNG50. {\bf Left:} a merger-free galaxy TNG50 475619, the total stellar mass of the satellite it ever accreted is smaller than $1\%$ of its current stellar mass. From top to bottom, the first panel is the dark matter (black dots), gas (blue dots), and stellar mass (red dots) of the main progenitor galaxy, and stellar mass assembly history of accreted satellites (red lines) before merged into the main progenitor; the second panel is the SFR (black) and sSFR (red) as a function of cosmic time; the third panel is the evolution of circularity distribution for stellar particles in the main progenitor galaxy, within the radius $r< 2.5$ kpc; the fourth panel is that for stellar particles at $r> 2.5$ kpc; the bottom panel is the stellar mass of disk, bulge, and hot inner stellar halo of the main progenitor galaxy. 
{\bf Right:} similar figures for a strongly stripped ETG TNG50 184937, it has had two massive mergers: a $1:1$ major merger at $z\sim 0.9$ (marked by the red vertical dashed line) and a $1:10$ merger at $z\sim 0.5$ (marked by the magenta vertical dashed line) and several other minor mergers. 
}
\label{fig:case}
\end{figure*}

\begin{figure*}
  \centering\includegraphics[width=14cm]{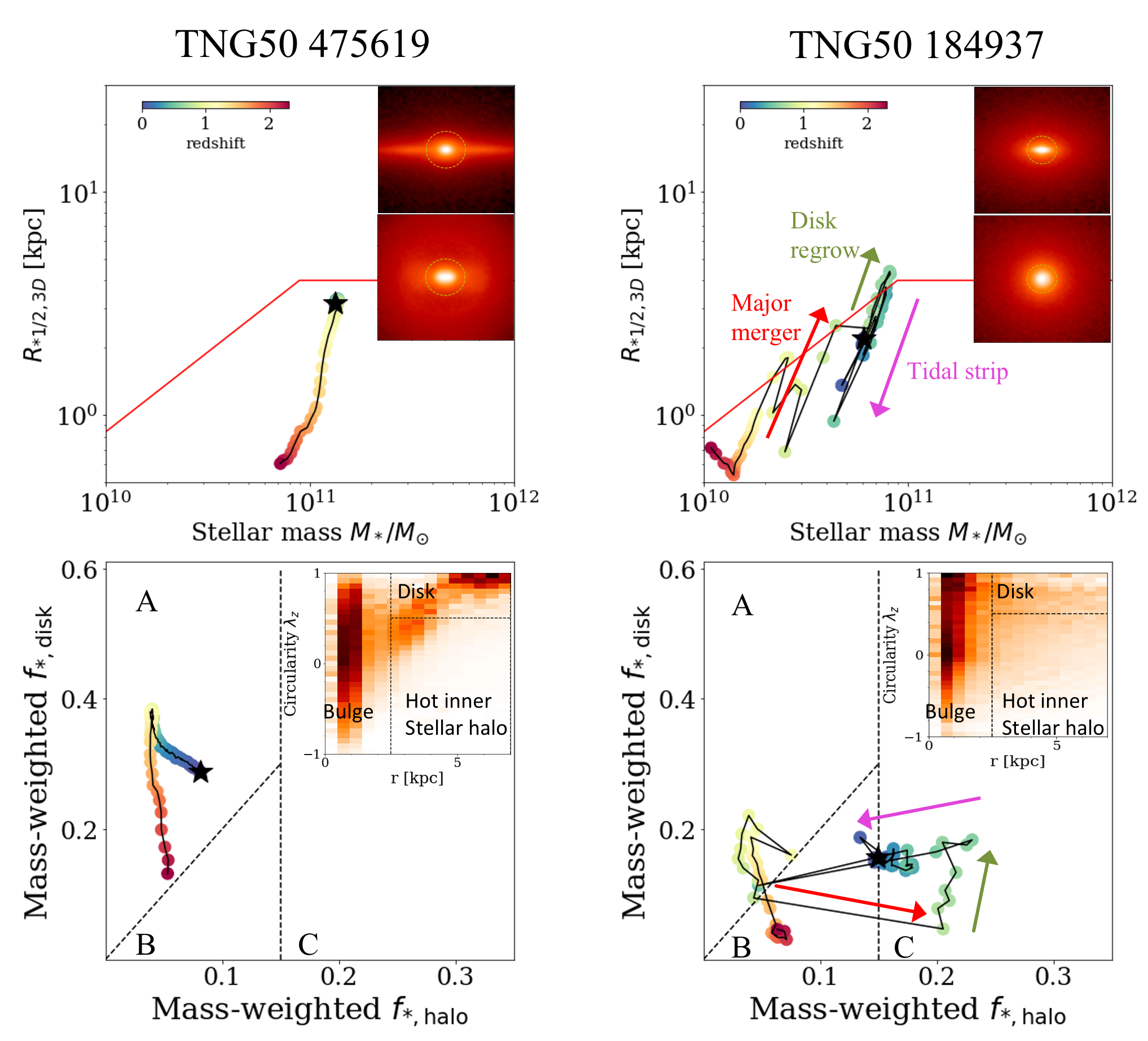}
\caption{The evolution of two typical compact ETGs, a merger-free galaxy TNG50 475619 in the left and a strongly stripped ETG TNG50 184937 in the right. {\bf Top:} evolution of the main progenitor galaxy in the stellar mass versus half-mass radius, coloured by the redshift as indicated by the colorbar, the star indicates its position at $z=0$, the insert panels are the edge-on and face-on surface brightness of the galaxy at $z=0$. {\bf Bottom:} evolution of the main progenitor galaxy in the mass fraction of the hot inner stellar halo $f_{*,\rm halo}$ versus disk $f_{*,\rm disk}$, the insert panels are the stellar orbit distribution $p(r, \lambda_z)$ of the galaxy at $z=0$. In the right panels for TNG50 184937, the red arrow points to the direction of evolution by a major merger, the green arrow indicates the regrowth of a disk, and the magenta arrow indicates tidal stripping.
}
\label{fig:case_MRe}
\end{figure*}

All the compact ETGs we selected from TNG50 are old with halted star formation.
The two major physical processes shaping these galaxies are accretion/merger and tidal stripping. Here we illustrate the formation of compact ETGs with two typical cases: one true merger-free galaxy with low accreted stellar mass fraction and no stripping (TNG50 475619), one stripped ETG experienced major merger and strong tidal stripping (TNG50 184973). We show their assembly history in Fig.~\ref{fig:case} and their structure evolution in Fig.~\ref{fig:case_MRe}. 

As shown in the left panel of Fig.~\ref{fig:case}, TNG50 475619 is a central galaxy with a quiescent merger history, the total stellar mass of satellite galaxies that it even accreted is about $1\%$ of its current stellar mass. It had a rapid star formation period at $z\sim 4-1.5$, which built most of its stellar mass. At $z\sim 1.5$, its gas mass gradually decreased and star formation stopped (sSFR $<10^{-11} M_*/yr^{-1}$) at $z\sim 0.7$, its total stellar mass remained unchanged since then. 
The stars were mostly born in dynamically cold orbits with $\lambda_z>0.5$, they
were then gradually heated to hot orbits with $\lambda_z \sim 0$ by a secular process at the inner 2.5 kpc, whereas the circularity distribution of stars at the outer regions was kept almost unchanged. Such secular heating of stars in the galaxy inner regions is common in TNG50 galaxies with quiescent merger histories, we refer to \citep{Zhang2025arXiv250107151Z} for a detailed discussion of it.

We see from Fig.~\ref{fig:case_MRe} that its mass and size grow gradually, with a steep increase in size since $z\sim 1.5$ with the growth of the disk; then it remains unchanged since $z\sim 1$ when the star formation rate becomes very low. 
As a result of quiescent merger history, TNG50 475619 maintains a very low fraction of halo, while the disk grows gradually with the new star formation, so it moves up in the plane of $f_{*,\rm halo}$ versus $f_{*,\rm disk}$ before $z\sim 1$, it stayed unchanged for a while, then slightly moves to the right-lower direction due to secure heating. 
We divide the plane into regions of A, B, and C, the same as those defined in Section~\ref{ss:fdiskfhalo}. TNG50 475619 stayed in region A for a whole life.

In contrast, TNG50 184937 is a galaxy that experienced several massive mergers and strong tidal stripping. It has a $\sim 1:1$ major merger at $z\sim 0.9$, which destroys the previous disk and causes a sudden increase of stars in the hot inner stellar halo, after the merger, the gas settled down, and new stars reform the disk. 
It has a second significant merger with a mass ratio of $\sim 1:10$ at $z\sim 0.5$. There are several peaks of star formation during the two mergers, and star formation is quickly quenched after the $\sim 1:10$ merger. After that the galaxy passed twice the pericenter of its orbits around a cluster center, both of its disk and hot inner stellar halo are partially stripped. 

TNG50 184937 shows a violent assembly history and, as a result, it has a complicated evolution in the mass-size plane.
It has a rapid increase in mass and size caused by the 1:1 major mergers at $z\sim 0.9$, which we illustrate by the red arrow in the right panel of Fig.~\ref{fig:case_MRe}; then it has a regrown disk, thus a steady increase in mass and size for a while (green arrow), followed by a decrease in mass and size due to tidal stripping since $z\sim 0.5$. 
Its location in the plane of $f_{*,\rm halo}$ versus $f_{*,\rm disk}$ also changes dramatically. It starts in region B and gradually grows to region A by inside-out growth; then it jumps to region C by the major merger at $z\sim 0.9$, which dramatically increases $f_{*,\rm halo}$ (red arrow); then it has a disk regrown after the merger (green arrow); followed by tidal stripping of both the disk and hot inner stellar halo, thus moving toward region B (magenta arrow). It could eventually reside in region B with more stars stripped.

In the end, these two example galaxies have had very different assembly histories located in similar regions in the mass-size plane; both are compact ETGs. However, they are located in different regions in the plane of $f_{*,\rm halo}$ versus $f_{*,\rm disk}$, where they can be well separated.

\section{Discussion}
\label{s:dis}
\subsection{The effects of observational data coverage}
Our analysis on TNG50 galaxies suggests that a compact ETG is guaranteed to be a true merger-free galaxy if it has a low fraction of hot inner stellar halo (equivalently small spheroid size) and a significant high disk fraction. Seven of the 15 compact ETGs from \citet{Y17} are thus taken as true merger-free galaxies that passed this criteria. For the remaining galaxies, we cannot provide a decisive conclusion on their accreted stellar mass fraction $f_{*,\rm accreted}$ based solely on their internal structural components. MRK 1216, NGC 1270, NGC 1271, and NGC 1277 are commonly taken as candidates of relic galaxies with their over-massive BH mass measured. NGC 1270 has a high fraction of hot inner stellar halo $f_{*, \rm halo} \sim 0.3$, its structure should be modified either by a relatively high accreted stellar mass fraction $f_{*,\rm accreted}$, or by tidal force or other physical interaction with the environment. MRK 1216 and NGC 1271 are still consistent with low accreted stellar mass fractions, but they are not guaranteed according to their current internal structure. NGC 1277 is among the seven guaranteed merger-free galaxies with a low fraction of hot inner stellar halo and a high fraction of disk. Thus, we survive one more criterium for being considered a true relic galaxy.

The structures are uniformly defined within 7 kpc for the simulated galaxies, while
there are a few cases with kinematic data coverage smaller than 7 kpc in the observed galaxies. For NGC 1277 ($R_e = 1.3$ kpc), PGC 12562 ($R_e = 0.7$ kpc), we have a data coverage of $r_{\rm max} \simeq 5$ kpc, which corresponds to $\sim 4 R_e$ and $\sim 7 R_e$ for the two galaxies, respectively. Thus, more than $95\%$ of the light should already be included. Importantly, they already appear disk-dominated within the data coverage and have a low fraction of stars in the hot inner stellar halo at $2.5<r<r_{\rm max}$, we expect a slight increase of $f_{*,\rm disk}$, and no significant increase of $f_{*,\rm halo}$ if extending to a larger radius. Their location in region A will not change. For NGC 1281 ($R_e = 1.6$ kpc) with $r_{\rm max} \sim 6$ kpc, there is already a relatively high fraction of stars in the hot inner stellar halo $f_{*,\rm halo}$ = 0.22 at $2.5<r<r_{\rm max}$, and we do not expect significant change of its location in region C with greater data coverage.

For NGC 1282 ($R_e = 1.3$ kpc), we have $r_{\rm max} \sim 4$ kpc, which also covers $3 R_e$ of the galaxy, but only a smaller region ($2.5<r<r_{\rm max}$) allowed us to define $f_{*,\rm disk}$ and $f_{*,\rm halo}$. There might be a relatively large uncertainty on $f_{*,\rm disk}$ and $f_{*,\rm halo}$ with the stellar orbit distribution constrained by the current data. We thus do not make a strong conclusion about this galaxy. A summary of the merger histories of all galaxies is included in table~\ref{tab:15compact}.

\subsection{Classic bulges formed without mergers}

The seven galaxies, NGC1277, PGC11179, UGC3816, NGC2767, PGC32873, PGC12562, and PGC70520, which we identified as merger-free galaxies, have massive compact bulges. Their bulges are dynamically hot and have old stellar populations, consistent with the common definition of a classic bulge.
In simulations, classic bulges can be formed without major mergers; for example, formed from secular heating in TNG50 (Zhang et al. 2024) or formed in dynamically hot gas in FIRE \citep{Yu2021}. The bulges in these seven galaxies are consistent with that born from secular heating in TNG50.

For most of them, we do not have evidence that they are merger-free
rigorously independent of the simulation predictions.
However, NGC 1277, served as a prototype merger-free relic galaxy, has independent evidence of a quiescent merger history from its single population of globular clusters \citep{Beasley2018}. 
NGC 1277 may serve as the first showcase of a galaxy with a classic bulge not formed by mergers.

\subsection{Identifying relic galaxies combing stellar population and stellar kinematics}
\label{SS:compare}

There are a few candidates in the local Universe, NGC 1277 \citep{Trujillo2014}, Mrk 1216 and PGC 032873 \citep{Ferr2017MNRAS.467.1929F} that are regarded as true relic galaxies based on their compact size, old stellar population, and short star formation timescales. These galaxies are also characterized by fast rotation and high central velocity dispersions. With a larger sample size, the Degree of Relicness (DoR) was introduced for massive compact galaxies in the INSPIRE project \citep{Spiniello2024MNRAS.527.8793S}, defined according to their star formation histories. Galaxies with the highest DoR are those that formed nearly all of their stars by $z=2$.

However, star formation history alone cannot reveal whether the structure of a galaxy has been altered by external dynamical processes, such as mergers or tidal stripping. This information is encoded in the structural properties of the galaxy, but previous studies have primarily relied on compactness as the key structural indicator, while stellar kinematics did not play a critical role in quantifying the relicness. In this work, we address this gap by carefully analyzing the internal dynamical structures of compact massive ETGs and comparing them between observations and simulations.

We find that a galaxy can be confidently identified as merger-free if it exhibits a low fraction of the hot inner stellar halo (or equivalently, an extremely compact spheroid) while retaining a dynamically cold disk. This finding is consistent with previous studies of compact ETGs from simulations \citep{Moura2024, Flores2022}, and aligns with observations of galaxies at high redshifts. The majority of massive galaxies observed at $z\sim 2$ are disk-dominated \citep{vanderWel2011, Kartaltepe2023} and likely fast-rotating \citep{DEugenio2024NatAs...8.1443D}. In general, massive galaxies at higher redshifts exhibit stronger rotation compared to massive galaxies at $z=0$ \citep{Bezanson2018, Zhang2025arXiv250107151Z}.

The combination of stellar population and stellar kinematic structures provides a comprehensive way towards understanding the relicness of compact massive ETGs. For the three well-studied relic candidates—NGC 1277 \citep{Trujillo2014}, Mrk 1216, and PGC 032873 \citep{Ferr2017MNRAS.467.1929F}—we confirm from the kinematics viewpoint that NGC 1277 and PGC 032873 are true merger-free relic galaxies with extremely quiescent merger histories (ex situ stellar mass fraction $< 5\%$). The internal structure of Mrk 1216 may have been moderately altered by subsequent mergers; it exhibits an ex situ stellar mass fraction of $\sim 10\%-20\%$ from the dynamical structures. It could still be defined as a relic galaxy if we take a less strict criterion of the ex-situ fractions for relics \citep{Quilis2013ApJ...773L...8Q}.

\section{Conclusion}
\label{s:conclusion}
We study the internal dynamical structures of 15 compact ETGs from observations and a comparable sample of 79 compact ETGs from TNG50. For the observed galaxies, we obtain the stellar orbit distribution by creating orbit-superposition models. These models are designed to fit the surface brightness and intrinsic luminosity density deprojected from the 2D image, as well as the kinematic maps from IFU observations. For the simulated galaxies, we obtain the stellar orbit distribution directly from the 6D phase-space information of the stellar particles. Based on the stellar orbit distribution, we dynamically decompose each galaxy into a disk, bulge, and hot inner stellar halo. We then characterize their internal structures by the luminosity fraction of each component. Alternatively, we dynamically define a hot spheroidal component and calculate its mass and half-light radius, which is qualitatively comparable to the photometrically decomposed bulge/spheroid.

For the compact ETGs selected from the central galaxies of TNG50, we find that the luminosity fraction of the hot inner stellar halo, or alternatively, the size of the hot spheroid, strongly correlates with the galaxies' accreted stellar mass fraction. Merger-free galaxies ($f_{\rm accreted}<0.1$) have extremely low fractions of hot inner stellar halo ($f_{*,\rm halo}<0.15$), or equivalently very compact hot spheroids ($R_{\rm Sph}<1$ kpc). However, for satellite galaxies, these features could also be the result of being heavily stripped by the tidal force. Tidal stripping will also destroy the dynamically cold disk (if one exists) when removing the outer spheroid component. Thus, a galaxy is guaranteed to be truly merger-free if it has a very low fraction of hot inner stellar halo ($f_{*,\rm halo}<0.15$) and a survived cold disk ($f_{*,\rm disk} > 2f_{*,\rm halo}$) for both central and satellite galaxies.

Compared directly with the dynamical structures of the compact ETGs of TNG50, we suggest that 7 of the 15 compact ETGs from observations \citep{Y17}, PGC 11179, UGC 3816, NGC 2767, NGC 1277, PGC 32873, PGC 12562, and PGC 70520, are true merger-free galaxies. The internal structures of UGC 2698 and NGC 1270 are highly likely to be altered by subsequent mergers or tidal forces. The rest galaxies (MRK 1216, NGC 0384, NGC 0472, NGC 1271, NGC 1281, and NGC 1282) are likely to still have relatively low frequency of mergers but are not guaranteed to be extremely merger-free.

We quantified the galaxy merger histories based on their internal orbital structures. This provides a critical way for identifying merger-free galaxies with extremely quiescent merger histories, rather than relying solely on galaxy mass and size. Combining stellar populations and our analysis on internal dynamical structures will help in identifying true relic galaxies and understanding the diversity of compact ETGs in many aspects, such as black hole mass, stellar populations, X-ray emission, DM distribution, and GC populations.

\begin{acknowledgement}
We thank Akın Yıldırım for providing us with the data of the 15 compact ETGs from observations, and we thank Annalisa Pillepich, Marie Martig, and Hans-Walter Rix for useful discussions.
The research presented here is partially supported by the CAS Project for Young Scientists in Basic Research under grant No. YSBR-062, and CAS PIFI project.
ACS acknowledges funding from CNPq and the Rio Grande do Sul Research Foundation (FAPERGS) through grants CNPq-11153/2018-6, PqG/FAPERGS-24/2551-0001548-5, FAPERGS/CAPES 19/2551-0000696-9. MTM acknowledges the Brazilian agency Conselho Nacional de Desenvolvimento Científico e Tecnológico (CNPq) through grant 140900/2021-7.
\end{acknowledgement}

\bibliographystyle{aa}  
\bibliography{aa53083-24} 

\end{document}